\documentclass[conference]{IEEEtran}
\IEEEoverridecommandlockouts
\usepackage{amsmath,amsfonts}
\usepackage{algorithmic}
\usepackage{graphicx}
\usepackage{textcomp}
\usepackage{xcolor}
\usepackage{array}
\usepackage[ruled, linesnumbered, vlined]{algorithm2e}
\usepackage{textcomp}
\usepackage{multirow}
\usepackage{bm}
\usepackage{booktabs}
\usepackage{ntheorem}
\usepackage{subfigure} 
\usepackage{epstopdf}
\usepackage{cases}

\newenvironment{sequation}{\begin{equation}\textstyle}{\end{equation}}

\usepackage{subfigure} 
\definecolor{color}{rgb}{0, 0, 0}
\definecolor{b}{rgb}{0, 0, 0}

\def\BibTeX{{\rm B\kern-.05em{\sc i\kern-.025em b}\kern-.08em
    T\kern-.1667em\lower.7ex\hbox{E}\kern-.125emX}}
\begin{document}
\title{IRS-assisted Edge Computing for Vehicular Networks: A Generative Diffusion Model-based Stackelberg Game Approach}

\author{
        \IEEEauthorblockN{Yixian Wang\IEEEauthorrefmark{2},
                          Geng Sun\IEEEauthorrefmark{2}\IEEEauthorrefmark{3}\IEEEauthorrefmark{1},
		               Zemin Sun\IEEEauthorrefmark{2}\IEEEauthorrefmark{1}, 
  		             Long He\IEEEauthorrefmark{2}, 
  		             Jiacheng Wang\IEEEauthorrefmark{3},
                          Shiwen Mao\IEEEauthorrefmark{4}
       }	
       \IEEEauthorblockA{
                        \IEEEauthorrefmark{2}College of Computer Science and Technology, Jilin University, Changchun 130012, China\\
                        \IEEEauthorrefmark{3}{School of Computer Science and Engineering, Nanyang Technological University, Singapore 639798, Singapore}\\
                        \IEEEauthorrefmark{4}Department of Electrical and Computer Engineering, Auburn University, Auburn 36849-5201, USA} 
                        E-mails: \{yixian23, helong23\}@mails.jlu.edu.cn, \{sungeng, sunzemin\}@jlu.edu.cn, jiacheng.wang@ntu.edu.sg, smao@ieee.org\\
	                  \IEEEauthorrefmark{1}Corresponding author: Geng Sun and Zemin Sun
        \vspace{-2em}
        }
\maketitle
\begin{abstract}
 Recent advancements in intelligent reflecting surfaces (IRS) and mobile edge computing (MEC) offer new opportunities to enhance the performance of vehicular networks. However, meeting the computation-intensive and latency-sensitive demands of vehicles remains challenging due to the energy constraints and dynamic environments. To address this issue, we study an IRS-assisted MEC architecture for vehicular networks. We formulate a multi-objective optimization problem aimed at minimizing the total task completion delay and total energy consumption by jointly optimizing task offloading, IRS phase shift vector, and computation resource allocation. Given the mixed-integer nonlinear programming (MINLP) and NP-hard nature of the problem, we propose a generative diffusion model (GDM)-based Stackelberg game (GDMSG) approach. Specifically, the problem is reformulated within a Stackelberg game framework, where generative GDM is integrated to capture complex dynamics to efficiently derive optimal solutions. Simulation results indicate that the proposed GDMSG achieves outstanding performance compared to the benchmark approaches.
\end{abstract}


\vspace{-0.5em}
\section{Introduction}
\label{sec_introduction}

\par With advancements in artificial intelligence (AI), vehicular networks are increasingly \textcolor{b}{leveraging AI} for emerging applications such as collision warning \cite{Fu2022} and autonomous driving \cite{Wang2019}. However, these applications generally require extensive computing resources and low latency, which pose a challenge given the limited processing power of vehicles. Mobile edge computing (MEC) \textcolor{b}{addresses} the abovementioned challenges by leveraging edge servers for task offloading and utilizing computational resources efficiently, thereby enhancing response times, energy efficiency, and overall system performance \cite{Porambage2018}. However, due to obstructions from dense buildings and frequent vehicle movement, the channel conditions continuously decay and vary, particularly in urban environments. These factors introduce complex interference to signal propagation, which results in a decline in the quality of vehicular communication links. 

\par To tackle this issue, intelligent reflecting surfaces (IRS) is identified as a promising technology \cite{Huang2023}. Specifically, IRS consists of numerous programmable reflecting elements that dynamically adjust signal reflections, thereby creating intelligent and adaptable wireless propagation environments. Research on IRS-assisted MEC has been growing to enhance vehicular network performance. For example, Li \textit{et al.} \cite{Li2024} investigated the mobility-aware offloading and resource allocation in non-orthogonal multiple access (NOMA) mobile edge computing to minimize the average task completion delay. Zhao \textit{et al.} \cite{Zhao2023} studied a symbiotic IRS-assisted MEC system, where edge users harvest radio frequency (RF) energy from a hybrid access point (HAP) and transfer tasks to \textcolor{b}{an} MEC server. The objective was to reduce the HAP energy usage through the joint optimization of offloading and beamforming strategies. However, these works primarily focused on either delay or energy consumption, while neglecting to consider both simultaneously, which may hinder comprehensive performance analysis. Furthermore, they often optimized only one or two aspects of task offloading, IRS phase shift-vector, and computation resource allocation, potentially limiting the effectiveness of the solution space. 

\par \textcolor{b}{Researchers have explored effective solutions to tackle} the complex optimization problems related to task offloading, IRS phase shift vector, and computation resource allocation. For example, Wan \textit{et al.} \cite{Wan2024} proposed a bisection search-based alternate feasibility verification sub-optimal algorithm to effectively tackle the joint optimization problem in IRS-assisted MEC systems. Moreover, Jiang \textit{et al.} \cite{Jiang2023} designed a deep reinforcement learning (DRL) approach that allows the agent to adjust optimization variables based on changing link conditions and randomly arriving task loads, thereby achieving higher performance. However, when dealing with complex search spaces, the bisection search algorithm often struggles to quickly adjust its strategy, thus resulting in limited flexibility, especially \textcolor{b}{in a vehicular network}. Additionally, DRL may encounter convergence issues and require longer training times in large-scale scenarios.

\par To address these challenges, there is growing interest in the generative diffusion model (GDM), which \textcolor{b}{offers} a novel perspective on optimization by effectively modeling complex distributions and adapting to varying conditions \cite{Cao2024}. Consequently, in this study, we explore the integration of GDM with Stackelberg game to solve the multi-objective optimization problem in IRS-assisted MEC for vehicular networks. The main contributions are summarized as follows:

\begin{itemize}
      \item We consider an IRS-assisted MEC architecture for vehicular networks. Specifically, the vehicles can offload delay-sensitive and computation-intensive tasks to the base station (BS), which enables rapid data processing and real-time decision support. Moreover, the IRS enhances the channel quality between vehicles and BS by intelligently adjusting the orientation of signal reflections, thereby enhancing the reliability and efficiency of data transmission. 
      
       \item We formulate a multi-objective optimization problem by considering the delay-sensitive requirements of vehicles and the energy limitations of \textcolor{b}{the} BS. Specifically, this problem aims to minimize the total task completion delay and total energy consumption of the system by optimizing the decisions of task offloading, IRS phase shift vector, and computation resource allocation. Moreover, we show that this problem is a mixed-integer non-linear programming (MINLP) and NP-hard problem.
      
      \item We propose a GDM-based Stackelberg game (GDMSG) approach to solve the problem. First, we reformulate the multi-objective optimization problem within a Stackelberg game framework to clarify the roles of decision-makers and simplify the problem. Furthermore, to capture the intricate dynamic characteristics and uncertainties of decision variables, we \textcolor{b}{incorporate} GDM within the framework to effectively derive the optimal solution.
	
      \item Simulations are conducted and the results indicate that the proposed GDMSG achieves outstanding performance with respect to total task completion delay, quality of experience (QoE) of vehicles, and revenue of BS compared with several benchmark approaches. Moreover, we find that the GDMSG approach exhibits better scalability in the considered scenario.  
\end{itemize}
\par The rest of this paper is structured as follows. Section \ref{sec_model} presents the system model and problem formulation. Section \ref{sec_GDM-based Stackelberg Game} introduces the GDMSG approach. Section \ref{sec_Simulation} \textcolor{b}{discusses} the simulation results. Finally, Section \ref{sec_conclusion} summarizes the paper.
\vspace{-1.5ex}
\section{System Model and Problem Formulation}
\label{sec_model}
\par In this section, an IRS-assisted MEC architecture is first introduced, followed by communication and computation models. Then, we formulate the joint optimization problem.
\vspace{-1.5ex}
\subsection{System Model}
\label{sec_system_model}
\vspace{-0.5ex}
{\color{color}
\par An IRS-assisted MEC architecture for vehicular networks is presented in Fig. \ref{fig_System_Model}, \textcolor{b}{with} $V$ vehicles, denoted as {$\mathcal{V}=\{1, \ldots, i, \ldots, V\}$}, and a BS with $S$ antennas \textcolor{b}{providing} services for single antenna vehicle $i$. Moreover, a building is equipped with a transparent dynamic IRS composed of $K$ reflective elements, which contains a controller for \textcolor{b}{adjusting} the phase shift of each reflecting element. The continuous system timeline is evenly divided into $N$ time slots with equal time duration $\delta_t$. Furthermore, in each time slot, each vehicle characterized by its speed and acceleration, could randomly generate at most one task that can be \textcolor{b}{either} carried out locally or offloaded to the BS via frequency division multiple access (FDMA). The task of vehicle $i$ is denoted as $U_i[n]=(D_i[n], {C_i[n]}, T_{i}^{\max})$, where $D_i[n]$ is the size (in bits), ${C_i[n]}$ is the computation intensity, and $T_{i}^{\max}$ is the deadline.
\begin{figure}[!t]
    \setlength{\abovecaptionskip}{0pt}
    \setlength{\belowcaptionskip}{0pt}
    \centering
   \includegraphics[width=3in]{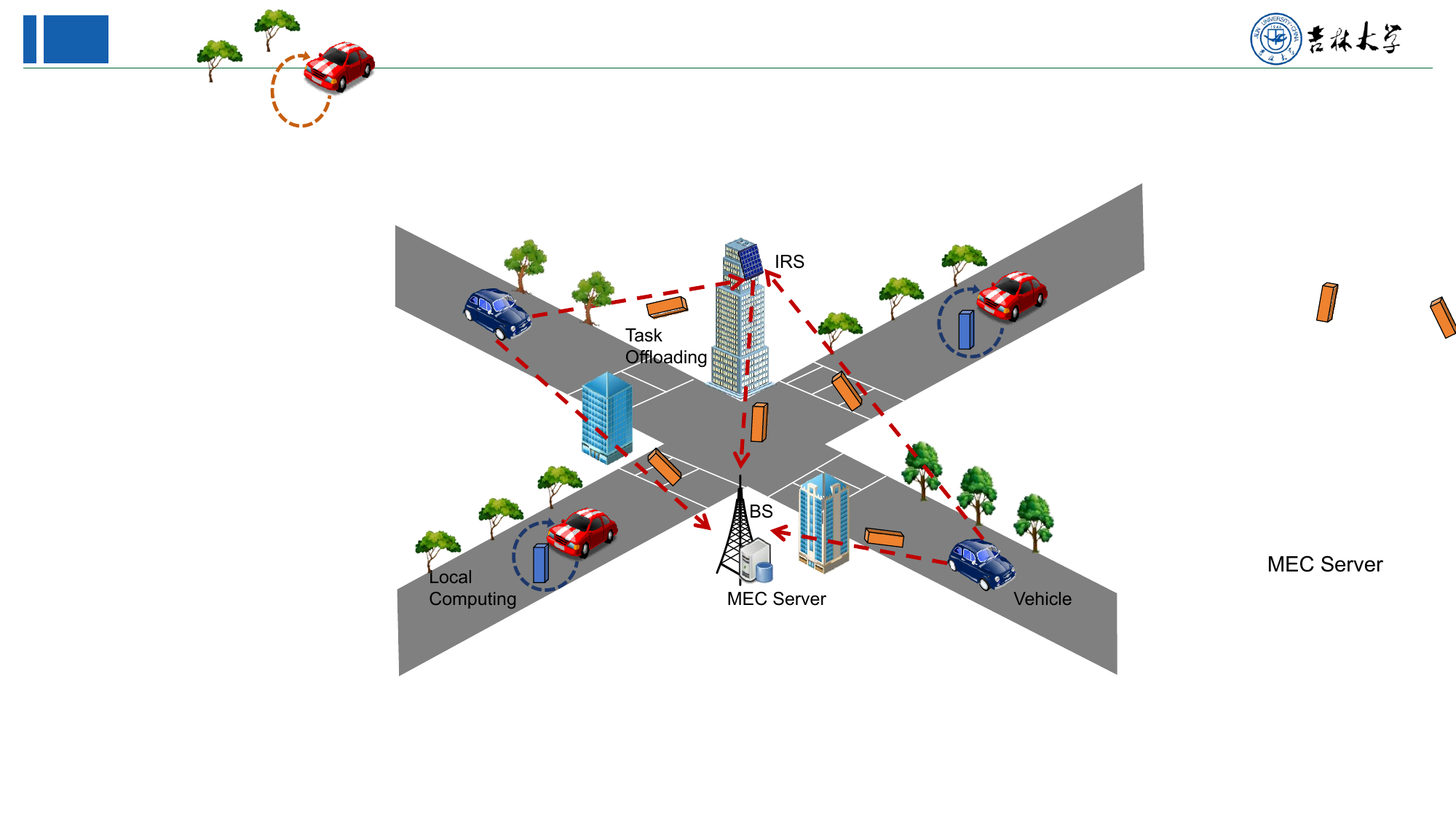}
    \caption{IRS-assisted MEC architecture for vehicular networks.}
    \label{fig_System_Model}
    \vspace{-3.5ex}
\end{figure}
\subsection{Communication Model}
\label{sec_communicationModel}
\par This work emphasizes the task uploading from vehicles to \textcolor{b}{the} BS while omitting the downloading, as computation outputs are typically much smaller than inputs. Moreover, FDMA is adopted for task uploading, as it divides the spectrum into subcarriers, with each vehicle transmitting on its assigned subcarrier \cite{Ali2016}.

\par Assuming that the BS has complete knowledge of the channel state information (CSI) for all relevant channels, we can model the channel gain from vehicle $i$ to the BS as Rayleigh fading, which is given as:
\vspace{-1ex}
\begin{equation}
\label{eq_i,b}
    \boldsymbol{h}_{i, b}[n]=\sqrt{\rho d_{i, b}^{-\alpha_{i, b}}[n]} \tilde {\boldsymbol{h}}_{i, b}[n],
\vspace{-1ex}
\end{equation}
\noindent where $\alpha_{i, b}$ is the path loss exponent, $\rho$ is the path loss at the reference distance $d_0 = 1 \mathrm{m}$, $d_{i, b}$ represents the distance between the vehicle and the BS, and $\tilde{\boldsymbol{h}}_{i, b}$ denotes a random scattering component following a Gaussian distribution \cite{Zhang2021}.

\par Furthermore, when buildings obstruct the propagation path between vehicle $i$ and the BS, communication signals experience significant penetration loss, which results in a \textcolor{b}{sharp} decrease in received signal strength. To address this issue, IRS can utilize intelligent beamforming techniques to adjust the angle and magnitude of reflected signals, thereby directing signals around buildings directly to the receiving end. Thus, this adjustment enhances the efficiency of signal transmission and improves the stability of communication.
\par The channel gain $\boldsymbol{h}_{i, r} \in \mathbb{C}^{1 \times K}$ from vehicle $i$ to the IRS follows the Rician distribution \cite{Zhang2017}, which is given as:
\begin{sequation}
\label{eq_i,r}
   \boldsymbol{h}_{i, r}[n]=\sqrt{\rho d_{i, r}^{-\alpha_{i, r}}[n]}\big(\sqrt{\frac{\gamma_{i, r}}{1+\gamma_{i, r}}} \boldsymbol{h}_{i, r}^{\mathrm{LoS}}[n]+\sqrt{\frac{1}{1+\gamma_{i, r}}} \boldsymbol{h}_{i, r}^{\mathrm{NLoS}}[n]\big),
\end{sequation}

\noindent where $\alpha_{i, r}$ is the associated path loss exponent, $\gamma_{i, r}$ is the Rician factor, and $d_{i, r}$ is the distance between the vehicle and the reference point of the IRS. Moreover, the line-of-sight (LoS) component and the non-line-of-sight (NLoS) component for the channel are denoted by $\boldsymbol{h}_{i, r}^{\mathrm{LoS}}$ and $\boldsymbol{h}_{i, r}^{\mathrm{NLoS}}$.

\par Similar to \cite{Di2020}, the channel gain $\boldsymbol{h}_{r, b} \in \mathbb{C}^{K \times S}$ from the IRS to the BS also follows the Rician distribution \cite{Zhang2017}. Moreover, the specific expression for $\boldsymbol{h}_{r, b}$ resembles that of \eqref{eq_i,r} and is omitted here due to space constraints.

\par In addition, the IRS reflection-coefficient matrix is given as:
\begin{equation}
\label{eq_matrix}
\boldsymbol{\Theta}=\operatorname{diag}(\beta e^{\iota \theta_1}, \ldots, \beta e^{\iota \theta_{K}}),
\end{equation}
\noindent where $\beta$ is set to 1 and $\iota$ is the imaginary unit ($\iota=\sqrt{-1}$). 
\par When vehicle $i$ chooses to offload its task via the IRS (i.e., $O_{i}^{o}[n]=1$), the received signal at the BS is given as:
\vspace{-1ex}
\begin{equation}
\label{eq_signal}
\boldsymbol{y}[n]=\sqrt{p_i^{t r}}[n](\boldsymbol{h}_{i,b}[n]+\boldsymbol{h}_{i,r}[n]\boldsymbol{\Theta} \boldsymbol{h}_{r,b}[n]) s_i[n]+\boldsymbol{z}_i,
\vspace{-1ex}
\end{equation}
\noindent where $p_i^{tr}$ and $s_i$ are the transmission power and unit-power signal of vehicle $i$, and $\boldsymbol{z}_i \sim \mathcal{C Z}\left(\mathbf{0}_{M \times 1}, \sigma^2 \boldsymbol{I}_M\right)$ represents the complex additive white Gaussian noise vector. The corresponding SINR, denoted by $\delta_{i,b}[n]$ is given as:
\vspace{-1ex}
\begin{equation}
\label{eq_SINR}
\delta_{i,b}[n]=\frac{p_i^{t r}\left\|\left(\boldsymbol{h}_{i,b}[n]+\boldsymbol{h}_{i,r}[n]\boldsymbol{\Theta} \boldsymbol{h}_{r,b}[n]\right)\right\|^2}{\sum_{j=1, j \neq i}^K p_j^{t r}\left\|\left(\boldsymbol{h}_{i, j}[n]+\boldsymbol{h}_{i,r}[n] \boldsymbol{\Theta} \boldsymbol{h}_{r, j}[n]\right)\right\|^2+\sigma^2}.
\end{equation}
\par Thus, the data transmission rate from vehicle $i$ to the BS by using FDMA is given as:
\vspace{-1ex}
\begin{equation}
\label{eq_rate}
    R_{i,b}[n]={\Omega}/{V}\log_2\left(1+\delta_{i,b}[n]\right),
\vspace{-1ex}
\end{equation}
\noindent where $\Omega$ is the system bandwidth.
\vspace{-1ex}
\subsection{Computation Model}
\label{sec_computation model}
\subsubsection{Service Delay} For task $U_i[n]$, the service delay for task completion is determined by the offloading strategy $O_i^a[n]$.

\par \textbf{\textit{Local Computing.}} The delay of vehicle $i$ to process task $U_i[n]$ locally is given as:
\vspace{-1ex}
\begin{equation}
\label{eq_local_delay}
    T_i^l[n]={\mathcal{C}_i^{\mathrm{req}}[n]}/{f_{i}[n]},
\vspace{-1ex}
\end{equation}
\noindent where $f_{i}[n]$ represents the computation capacity of vehicle $i$.

\par \textbf{\textit{Offloading Computing.}} When task $U_i[n]$ is processed by the BS, the offloading delay primarily comprises transmission and computation delays, which is given as:
\vspace{-1ex}
\begin{equation}
\label{eq_BS_delay}
   T_i^o[n]={T_{i, b}^{\text {tran}}[n]}+{T_{b, i}^{\text {comp}}[n]},
\vspace{-1ex}
\end{equation}
\noindent \textcolor{b}{where} ${T_{i, b}^{\text {tran}}[n]}={D_i[n]}/{R_{i, b}[n]}$, $T_{b, i}^{\text {comp}}[n]={\mathcal{C}_i^{\mathrm{req}}[n]}/{f_{b, i}[n]}$, \textcolor{b}{and} $f_{b, i}[n]$ is the computation resources allocated by the BS to the task.

\par According to \eqref{eq_local_delay} and \eqref{eq_BS_delay}, the total task completion delay across $N$ slot is given as:
\vspace{-1ex}
\begin{equation}
\label{eq_delay_total}
T_{\mathrm{total}}=\sum_{n \in \mathcal{N}}\sum_{i \in \mathcal{V}}\left( T_{i}^{l}[n]+T_{i}^{o}[n]\right).
\vspace{-1ex}
\end{equation}

\subsubsection{Energy Consumption}
\label{sec_energy consumption}
\par Completing task $U_i[n]$ may result in additional costs for vehicles or BS.
\par \textbf{\textit{Local Computing.}} The energy consumption of vehicle $i$ to process task $U_i[n]$ locally is given as:
\vspace{-1ex}
\begin{equation}
	\label{eq_energyLocal}
        E_{i}^{l}[n]=\kappa_i(f_i[n])^2\mathcal{C}_i^{\mathrm{req}}[n],
\vspace{-1ex}
\end{equation}
\noindent where $\kappa_i\geq0$ is the effective switched capacitance of the CPU of vehicle $i$.

\par \textbf{\textit{Offloading Computing.}} When task $U_i[n]$ is processed by the BS, the energy consumption mainly includes transmission and computation energy consumptions, which is given as:
\vspace{-1ex}
\begin{equation}
\label{eq_BS_energy}
   E_i^o[n]={E_{i, b}^{\text {tran}}[n]}+{E_{b, i}^{\text {comp}}[n]},
\vspace{-1ex}
\end{equation}
\noindent \textcolor{b}{where} ${E_{i, b}^{\text {tran}}[n]}=$ {\small$p_i^{tr}[n]{T_{i, b}^{\text {tran}}[n]}$}, $E_{b,i}^{\text {comp}}[n]=$ {\small$\kappa_b(f_{b, i}[n])^2\mathcal{C}_i^{\mathrm{req}}[n]$}, \textcolor{b}{and} $\kappa_b\geq0$ denotes the effective switched capacitance of the CPU of BS.

\par According to \eqref{eq_energyLocal} and \eqref{eq_BS_energy}, the total energy consumption across $N$ time slots is given as:
\vspace{-1ex}
\begin{equation}
\label{eq_energy_total}
E_{\mathrm{total}}=\sum_{n \in \mathcal{N}}\sum_{i \in \mathcal{V}}\left( E_{i}^{l}[n]+E_{i}^{o}[n]\right).
\end{equation}
\vspace{-3ex}
\subsection{Problem Formulation}
\label{sec_Problem Formulation}
\vspace{-1ex}
\par This work aims to minimize the total task completion delay and total energy consumption of the system by jointly optimizing task offloading $\mathbf{O} = \{O_i^a[n]\}_{i \in \mathcal{V}, a \in \mathcal{A}, n \in \mathcal{N}}$, IRS phase shift vector $\boldsymbol{\theta} = \{\theta_k[n]\}_{k \in \mathcal{K}, n \in \mathcal{N}}$, and computation resource allocation $\mathbf{F} = \{f_{b, i}[n]\}_{i \in \mathcal{V}, n \in \mathcal{N}}$. Therefore, the problem is given as:
\vspace{-1.7ex}
\begin{subequations}
	\label{eq_problem}
	\begin{alignat}{2}
		\mathbf{P}: \quad & \min _{\mathbf{O},\boldsymbol{\theta},\mathbf{F}} \ \left\{T_{\mathrm{total}}, E_{\mathrm{total}}\right\}\label{p}\\
		\text{s.t.} \quad 
              &O_i^a[n]\in\{0,1\}, \forall i\in \mathcal{V}, a\in \mathcal{A}, n \in \mathcal{N}\label{1}\\
             &\sum_{a\in \mathcal{A}}O_i^a[n]\leq 1, \ \forall  i\in \mathcal{V}, n \in \mathcal{N}\label{2}\\
             &O_i^a[n]T_i^a[n]\leq T_i^{\max}, \ \forall  i\in \mathcal{V},a\in \mathcal{A}, n \in \mathcal{N}\label{3}\\
             &\zeta_i[n]\in\{0,1\},\forall i\in \mathcal{V}, n \in \mathcal{N}\label{4}\\
              &0 \leq f_{b,i}[n] \leq f_b^{\text{max}}, \ \forall i\in \mathcal{V}, n \in \mathcal{N}\label{5}\\
             &\sum_{i\in \mathcal{I}_0} f_{b,i}[n] \leq f_b^{\text{max}}, \forall  n \in \mathcal{N}\label{6}\\
             & 0 \leq \theta_k[n]<2 \pi,\ \forall k \in \mathcal{K}, n \in \mathcal{N}\label{7}
	\end{alignat}
\end{subequations}
\noindent Constraints \eqref{1}-\eqref{2} define the offloading strategies, which indicate that the vehicle can choose either local or offloading. Constraint \eqref{3} ensures task completion before the deadline. Constraint \eqref{4} represents each vehicle generates at most one task. Constraints \eqref{5} and \eqref{6} ensure that the BS allocates positive resources within its maximum limit. Moreover, constraint \eqref{7} specifies the phase shift of each IRS element $K$. 

\par The abovementioned problem $\mathbf{P}$ involves binary variables (i.e., task offloading $\mathbf{O}$) and continuous variables (i.e., IRS phase shift-vector $\boldsymbol{\theta}$ and computation resource allocation $\mathbf{F}$). Consequently, problem $\mathbf{P}$ is a mixed-integer non-linear programming (MINLP) problem, which is also non-convex and NP-hard. Therefore, it is difficult to find the optimal solution to problem $\mathbf{P}$, which motivates us to propose the GDMSG.
\vspace{-1.6ex}
\section{GDMSG for IRS-assisted MEC System}
\label{sec_GDM-based Stackelberg Game}
\vspace{-0.5ex}
\par In this section, the GDMSG is proposed to solve the formulated problem. First, we reformulate the multi-objective optimization problem within a Stackelberg game framework, which clarifies the roles of the participants and \textcolor{b}{reduces} the complexity of the problem. This reformulation is essential as it facilitates interaction between roles to effectively manage dependencies among different objectives, and optimize the decision-making process. Moreover, since GDM has the ability to effectively simulate complex real-world scenarios and capture dynamic characteristics, we employ it to derive more stable and efficient solutions for the Stackelberg game.
\vspace{-1.8ex}
\subsection{Stackelberg Game Formulation}
\label{sec_Stackelberg Game Formulation}
\vspace{-0.4ex}
\par In this subsection, we reformulate the formulated optimization problem \textcolor{b}{into} a Stackelberg game framework. 

\subsubsection{QoE of vehicles}
\par \textbf{Task completion revenue.} For each vehicle, the task completion revenue can be measured by the difference between the task deadline and the task completion delay \cite{Tong2023}. Thus, the revenue function of vehicle $i$ is given as:
\vspace{-1.8ex}
\begin{equation}
\label{eq_Task completion revenue}
U_i^{rev}[n]=\log \left(c+T_i^{\max }[n]-T_i[n]\right),
\vspace{-1ex}
\end{equation}
\noindent where $c$ is a positive value used to adjust the value of $U_i^{rev}[n]$ to a reasonable range.

\par \textbf{Task offloading cost.} For the task of each vehicle, the costs consist of the energy consumption and the fees paid to the server. Thus, the cost function of vehicle $i$ is given as:
\vspace{-1ex}
\begin{equation}
\label{eq_Task offloading cost}
    U_i^{cost}[n] = 
    {\begin{cases}
    E_{i}^{l}[n], & O_i^l[n]=1 \\
    E_{i}^{o}[n]+\left(G_i^{\max }-f_{b, i}[n] {\rho}_{b, i}[n]\right), & O_i^o[n]=1
\end{cases}},
\end{equation}
\noindent where ${\rho}_{b, i}[n]$ is the unit price of computing resources for vehicle $i$, and $G_i^{\max}$ represents the budget of vehicle $i$.

\par Therefore, the utility function of vehicle $i$ is given as:
\vspace{-1ex}
\begin{align} 
\label{eq_i,a} 
& U_{i,a}[n]=w_i \log \left(c+T_i^{\max }[n]-T_i[n]\right) -\left(1-w_i\right)  \\ \nonumber 
& \mathbb{I}_{(a=l)}{ E_{i}^{l}[n]} + \mathbb{I}_{(a=o)}({E_{i, b}^{\text {tran}}[n]}+{G_i^{\max }-f_{b, i}[n] {\rho}_{b, i}[n]}). 
\vspace{-1ex}
\end{align}

\subsubsection{Revenue of BS}
\par The utility of the BS for task $U_i[n]$ is the difference between the reward received from vehicle $i$ and the cost of energy consumption, which is given as:
\vspace{-1ex}
\begin{equation}
\label{eq_Revenue of BS}
U_{b,i}[n]=w_b {f_{b,i}[n] {\rho}_{b,i}[n]-{G}_{b}^{\min }}-\left(1-w_b\right) {E_{b, i}^{\text {comp}}[n]}, 
\vspace{-1ex}
\end{equation}
\noindent where ${G}_{b}^{\min }$ is the minimum price for the computing resources.

\par Accordingly, the Stackelberg game can be formulated by using the leader-follower model.
\par The follower game of the vehicle is formulated as:
\vspace{-1.5ex}
\begin{subequations}
	\label{eq_problem1}
	\begin{alignat}{2}
		&\max _{\mathbf{O},\boldsymbol{\theta}} \ U_{i,a}[n] \label{p1}\\
	    & \ \text{s.t.} \quad \eqref{1} \sim \eqref{4}, \eqref{7}. \label{1.1}
	\end{alignat}
\end{subequations}
\vspace{-3.5ex}
\par The leader game of the BS is formulated as:
\vspace{-1.5ex}
\begin{subequations}
	\label{eq_problem2}
	\begin{alignat}{2}
		&\max _{\mathbf{F}} \ U_{b,i}[n] \label{p2}\\
		& \ \text{s.t.} \quad \eqref{5}, \eqref{6}. \label{2.1}
	\end{alignat}
\end{subequations}
\vspace{-4ex}
\subsection{GDM-based Stackelberg Game Solution}
\label{sec_Generative Diffusion}
\vspace{-0.3ex}
\par In this subsection, the process of utilizing the GDM to find the optimal Stackelberg game solution is introduced.

\subsubsection{Modeling the IRS-assisted Vehicular Network Environment}
\par We first define the IRS-assisted vehicular network environment, which provides the foundation for the strategic interactions within the game. Specifically, to more effectively capture a diverse range of network conditions and strategic scenarios, the initial configuration of the state space is generated through random sampling.

\subsubsection{Configuring GDM with Stackelberg Game and Dynamic Reward Mechanism} 
\par We configure the GDM based on the Stackelberg game, where the action space includes all possible strategies for the BS, followed by the strategies for the vehicles. Moreover, we design a reward mechanism \textcolor{b}{that features a centralized controller at the BS} to effectively evaluate and monitor the performance of system. In particular, rewards are established by comparing the current state and the previous state. If the total utility $U^j$ including the QoE of vehicles and the revenue of BS increases compared to the previous round, the reward is based on the improvement and the corresponding weight. If there is no increase, the reward is set to zero to ensure that only the positive changes are rewarded. This mechanism encourages the GDM to adopt strategies that improve the total utility over time, which is as follows:
\vspace{-1ex}
\begin{equation}
\label{reward}
R^j = 
\begin{cases} 
U^j - U^{j-1}, & \text{if } U^j > U^{j-1}, \\
0, & \text{otherwise}, 
\end{cases}
\end{equation}
\noindent where $U_{\mathrm{QoE}}=\sum_{n \in \mathcal{N}}\sum_{i \in \mathcal{V}}\sum_{a \in \mathcal{A}} U_{i,a}[n]$, $U_{\mathrm{Revenue}}=\sum_{n \in \mathcal{N}}\sum_{i \in \mathcal{V}} U_{b,i}[n]$, $U^j = w_i U_{QoE}^j + w_b U_{Revenue}^j$.
\subsubsection{Training GDM with Noise and Denoising for Stackelberg Game} 
\par To effectively manage the complexities and uncertainties in strategy generation for the Stackelberg game, we employ the denoising diffusion probabilistic model (DDPM) \cite{ho2020denoising}, which operates in forward and reverse phases. In the forward phase, the model adds Gaussian noise until the data becomes a pure noise distribution, while in the reverse phase, the model removes the noise to reconstruct the original data.

\par \textbf{Forward Phase:} For a given initial strategy sample $\mathbf{x}_0$, the forward phase progressively adds Gaussian noise over $T$ steps to generate a series of noisy samples $\{\mathbf{x}_t \}_{t=1}^T$. Specifically, the distribution of each noisy sample is given as:
\vspace{-1ex}
\begin{equation}
\label{distribution}
q(\mathbf{x}_t | \mathbf{x}_{t-1}) = \mathcal{N}(\mathbf{x}_t; (1 - \beta_t) \mathbf{x}_{t-1}, \beta_t \mathbf{I}),
\vspace{-1ex}
\end{equation}
\noindent where $\beta_t=1-e^{-\beta_{\min}/{T}-{2 t-1}/{2 T^2}\bigl(\beta_{\max }-\beta_{\min }\bigl)}$ is a variance schedule, and $\mathbf{I}$ represents the identity matrix. 

\par Thus, the forward process from $\mathbf{x}_0$ to $\mathbf{x}_T$ is given as:
\vspace{-1.5ex}
\begin{equation}
\label{forward process}
q(\mathbf{x}_{1:T}|\mathbf{x}_0) = \prod_{t=1}^T q(\mathbf{x}_t|\mathbf{x}_{t-1}).
\vspace{-1ex}
\end{equation}

\par In addition, it is important to note that the forward process allows for sampling in closed form at any time step $t$. This property provides flexibility in generating noisy samples during the forward process, establishing a foundation for the subsequent reverse process. Specifically, the sample at any given moment $\mathbf{x}_t$ can be expressed as a combination of the initial data $\mathbf{x}_0$ and random noise, which is given as:
\vspace{-1ex}
\begin{equation}
\label{closed form}
\mathbf{x}_t=\sqrt{\bar{\alpha}_t} \mathbf{x}_0+\sqrt{1-\bar{\alpha}_t} \boldsymbol{\epsilon},
\vspace{-1ex}
\end{equation}
\noindent where $\epsilon\sim\mathcal{N}(0,\mathbf{I})$ represents Gaussian noise, and $\bar{\alpha}_t = \prod_{s=1}^{t}\alpha_s$ denotes the cumulative product of $\alpha_s$, where $\alpha_t=1-\beta_t$ describes the overall data retention after $t$ steps. However, note that due to the lack of an optimal decision strategy (i.e., $\mathbf{x}_0$ in the forward phase) for the problem, the forward phase is not integrated into the proposed GDMSG.

\par \textbf{Reverse Phase:} The reverse phase gradually removes noise from the sample $\mathbf{x}_T$ to recover the real data $\mathbf{x}_0$ through a series of transitions governed by Gaussian distributions. Each latent variable $\mathbf{x}_{t-1}$ is generated based only on the current latent variable $\mathbf{x}_t$, which is given as:
\vspace{-1ex}
\begin{equation}
\label{distribution_reverse}
p_\delta(\mathbf{x}_{t-1} \mid \mathbf{x}_t)=\mathcal{N}(\mathbf{x}_{t-1} ; \boldsymbol{\mu}_\delta(\mathbf{x}_t, t), \tilde{\beta}_t \mathbf{I}),
\vspace{-1ex}
\end{equation}
\noindent where $\boldsymbol{\mu}_\delta(\mathbf{x}_t, t)=\frac{\sqrt{\bar{\alpha}_{t-1}} \beta_t}{1-\bar{\alpha}_t} \mathbf{x}_0+\frac{\sqrt{\alpha_t}\left(1-\bar{\alpha}_{t-1}\right)}{1-\bar{\alpha}_t} \mathbf{x}_t$ and $\tilde{\beta}_t=\frac{1-\bar{\alpha}_{t-1}}{1-\bar{\alpha}_t} \beta_t$ are the mean and variance for the denoising model, and $\delta$ denotes its parameters. 

\par However, since the diffusion model can only access the current latent variable $\mathbf{x}_t $ during the reverse phase and cannot directly obtain the real data $\mathbf{x}_0 $, we need to reorganize the expression for $\boldsymbol{\mu}_\delta(\mathbf{x}_t, t)$. Specifically, according to \eqref{closed form}, we first express $\mathbf{x}_0 $ as a function of $\mathbf{x}_t $ and the predicted noise of the model, which is given as:
\vspace{-1ex}
\begin{equation}
\label{function}
\mathbf{x}_0={1}/{\sqrt{\bar{\alpha}_t}}(\mathbf{x}_t-\sqrt{1-\bar{\alpha}_t} \boldsymbol{\epsilon}_\delta(\mathbf{x}_t)),
\vspace{-1ex}
\end{equation}
\noindent where $\boldsymbol{\epsilon}_\delta$ represents a function approximator that predicts noise $\boldsymbol{\epsilon}$ based on current latent variable $\mathbf{x}_t$. Then, the parameterized mean function $\boldsymbol{\mu}_\delta(\mathbf{x}_t,t)$ is given as:
\vspace{-1ex}
\begin{equation}
\label{parameterized mean function}
\boldsymbol{\mu}_\delta(\mathbf{x}_t, t)={1}/{\sqrt{\alpha_t}}(\mathbf{x}_t-{\beta_t}/{\sqrt{1-\bar{\alpha}_t}} \boldsymbol{\epsilon}_\delta(\mathbf{x}_t, t)).
\vspace{-1ex}
\end{equation}
\par Thus, the reverse phase from $\mathbf{x}_t$ to $\mathbf{x}_0$ is given as:
\vspace{-1.5ex}
\begin{equation}
\label{reverse_process}
p_\delta\left(\mathbf{x}_{0: T}\right)=p\left(\mathbf{x}_T\right) \prod_{t=1}^T p_\delta\left(\mathbf{x}_{t-1} \mid \mathbf{x}_t\right).
\vspace{-1ex}
\end{equation}
\par Accordingly, the reverse phase is outlined in Algorithm~\ref{Algorithm Sampling}.
\vspace{-1.5ex}
\begin{algorithm}
    \caption{Sampling Algorithm for GDM}
    \label{Algorithm Sampling}
    \SetAlgoLined
    \KwIn{Steps $T$, diffusion model parameters $\delta$.}
    \KwOut{Generated sample $\mathbf{x}_0$.}
    Initialize a random Gaussian distribution $\mathbf{x}_T \sim \mathcal{N}(0, \mathbf{I})$; \\
    \For{$t = T$ \textbf{to} $1$}
    {
        Calculate the mean $\boldsymbol{\mu}_\delta(\mathbf{x}_t, t)$ based on the current latent variable and model parameters; \\
        Calculate the distribution $p(\mathbf{x}_{t-1} \mid \mathbf{x}_t, \delta)$; \\
        Sample $\mathbf{x}_{t-1} \sim \mathcal{N}(\boldsymbol{\mu}_\delta(\mathbf{x}_t, t), \tilde{\beta}_t \mathbf{I})$; \\      
    }
    \Return{$\mathbf{x}_0$}\;
\end{algorithm}
\vspace{-4ex} 
\subsection{Main Steps of GDMSG and Analysis}
\label{sec_Analysis of GDMSG}
\vspace{-0.5ex} 
\par \textcolor{b}{The} steps of GDMSG are presented in Algorithm \ref{Algorithm GDMSG}. First, the network state, reward, and convergence threshold are initialized (Lines 1 to 2). Then, in each iteration, the strategy of leader is determined using the sampling algorithm, followed by the acquisition of follower strategies for each vehicle (Lines 5 to 8). Subsequently, the utility for each vehicle and the revenue for the BS are calculated (Lines 9 to 10). Finally, the reward is updated, convergence is checked, and the optimal strategies are returned (Lines 12 to 19).
\vspace{-2ex} 
\begin{algorithm}
    \caption{The Proposed GDMSG Algorithm}
    \label{Algorithm GDMSG}
    \SetAlgoLined
    \KwIn{Time slots $N$, iterations $J_{\text{iter}}$}
    \KwOut{$\mathbf{O}^{opt}$, $\boldsymbol{\theta}^{opt}$, $\mathbf{F}^{opt}$}
    Initialize the network state through random sampling; \\
    Initialize reward $R^0 = 0$, threshold $\varepsilon = 10^{-2}$; \\

    \For{$j = 1$ \textbf{to} $J_{\text{iter}}$}
    {
        Call Algorithm~\ref{Algorithm Sampling} to obtain leader strategy $\mathbf{F}^j$ from $\mathbf{x}_0$; \\

        \For{$n = 1$ \textbf{to} $N$}
        {
            \For{$i = 1$ \textbf{to} $V$}
            {
                Obtain follower strategy $\boldsymbol{\theta}^j$ based on leader strategy $\mathbf{F}^j$; \\

                Obtain IRS phase shift strategy $\mathbf{O}^j$ based on $\mathbf{F}^j$ and $\boldsymbol{\theta}^j$; \\
                
                Calculate $U_{i,a}[n]$ based on Eq.~\eqref{eq_i,a}; \\
                Calculate $U_{b,i}[n]$ based on Eq.~\eqref{eq_Revenue of BS}; \\
            }
        }
        Calculate $R^j$ based on $\mathbf{a}$ using Eq.~\eqref{reward}; \\

        Update $R^{j-1} \gets R^j$; \\

        \If{$|R^j - R^{j-1}| < \varepsilon$}
        {
            \textbf{break};  
        }
    }
    \Return{$\left\{ \mathbf{O}^{opt}, \boldsymbol{\theta}^{opt}, \mathbf{F}^{opt} \right\}$};
\end{algorithm}
\vspace{-2ex} 
\par The complexity of the GDMSG algorithm is $O(J_{\text{iter}}NVT)$, where $J_{\text{iter}}$, $N$, and $V$ represent the number of iterations, time slots, and vehicles, respectively. Moreover, $T$ is the complexity of the sampling algorithm, as the sampling process follows a reverse diffusion process over $T$ steps, with each step involving constant-time operations.

\begin{figure*}[!hbt] 
	\centering
	\subfigure[]
	{
		\begin{minipage}[t]{0.23\linewidth}
			\centering
		\includegraphics[scale=0.3]{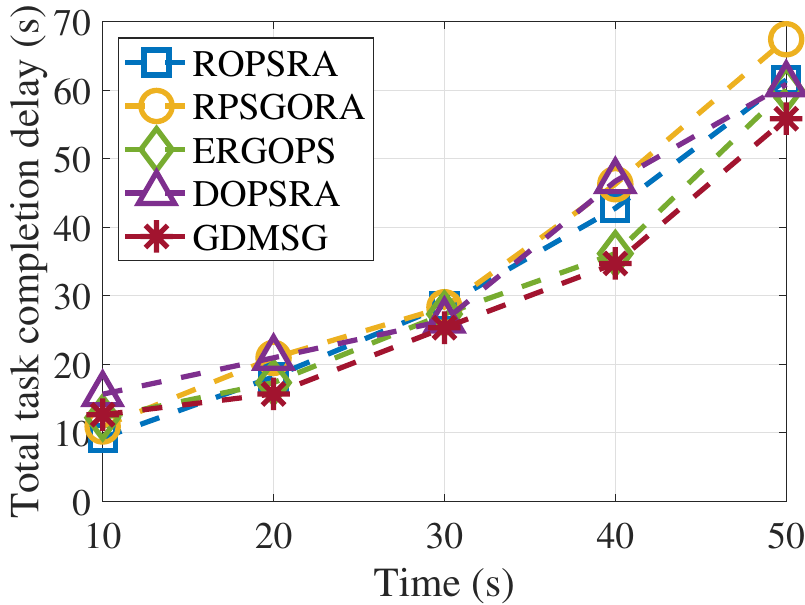}
		\end{minipage}
	}
	\subfigure[]
	{
		\begin{minipage}[t]{0.23\linewidth}
			\centering
		\includegraphics[scale=0.3]{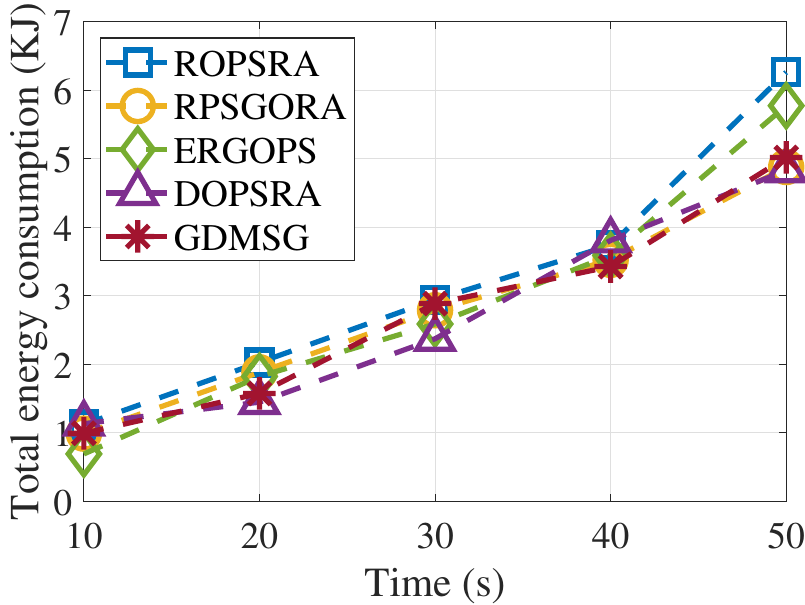}	
		\end{minipage}
	}
	\subfigure[]
	{
		\begin{minipage}[t]{0.23\linewidth}
			\centering
		\includegraphics[scale=0.3]{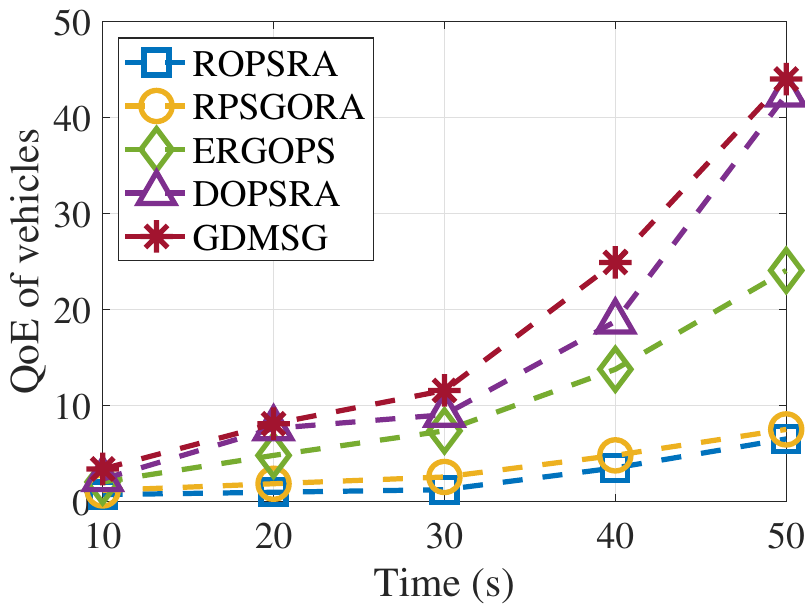}
		\end{minipage}
	}
        \subfigure[]
	{
		\begin{minipage}[t]{0.23\linewidth}
			\centering
		\includegraphics[scale=0.3]{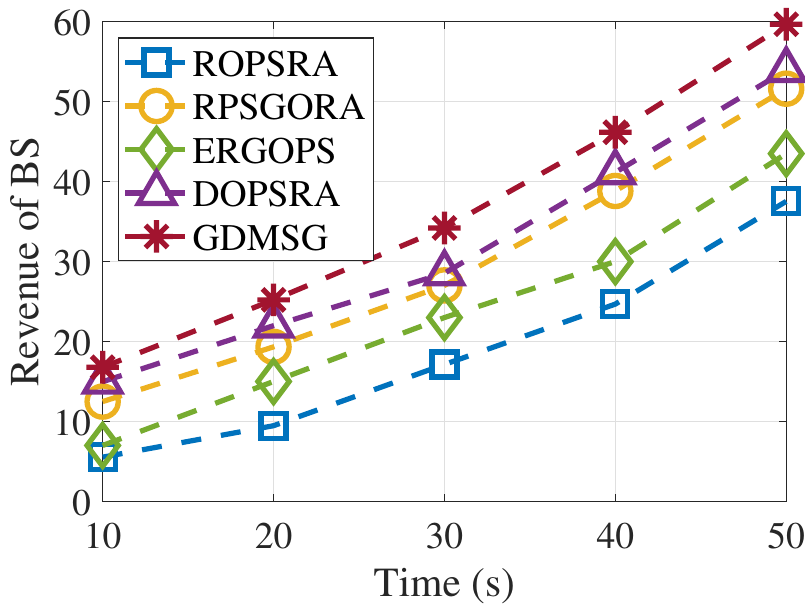}
		\end{minipage}
	}
	\centering
\vspace{-0.6em}
	\caption{System performance with time. (a) Total task completion delay. (b) Total energy consumption. (c) QoE of vehicles. (d) Revenue of BS.}
	\label{fig_time}
\vspace{-1.5em}
\end{figure*}
\vspace{-1ex}
\section{Simulation Results and Analysis}
\label{sec_Simulation}
\vspace{-0.5ex}
\par In this section, simulation results are presented to validate the effectiveness of the GDMSG approach.
\vspace{-1ex}
\subsection{Simulation Setup}
\label{sec_Simulation_Setup}
\vspace{-0.2ex}
\par \textbf{Scenarios.} We study an IRS-assisted MEC architecture for vehicular networks. Specifically, a BS is deployed at a fixed location to offer offloading services for $6$ vehicles, which are randomly placed within a $100 \times 100 \ \text{m}^2$ area.
\par \textbf{Parameters.} For the vehicles, the parameters are set as follows: computing capability $f_i$ = 1 GHz, transmission power $p_i^{tr} = 30$ dBm, the effective switching capacitance $\kappa_i$ = $10^{-27}$, and the deadline of the task $T_{i}^{\max}$ = [0.1, 10]s. For the BS, the parameters are set as follows: computing capability $f_{b,i}$ = 5 GHz, the effective switching capacitance $\kappa_b$ = $10^{-26}$, and the three-dimensional coordinates is (80 m, 40 m, 0). For the IRS, the three-dimensional coordinates is (0, 0, 50 m). \textcolor{b}{Moreover}, the communication parameters are set as follows: the path loss $\rho$ = -20 dB, the channel bandwidth $\Omega$ = $10$ MHz, and the noise power $\sigma^2$ = -114 dBm. Moreover, the Rician factor is conﬁgured as 3 dB, and the path loss exponents for BS-vehicles, BS-IRS, and IRS-vehicles links are 3, 2.2 and 2.5, respectively.

\par \textbf{Benchmarks.} This work conducts an evaluation of the proposed GDMSG compared to the following approaches.
\begin{itemize}
	\item  \textit{Random offloading, phase shift, and resource allocation (ROPSRA)}: the task offloading, IRS phase shift vector, and computation resource allocation strategies are decided randomly.
	\item  \textit{Random phase shift, GDMSG-based offloading and resource allocation (RPSGORA)}: the IRS phase shift vector strategies of vehicles are decided randomly, while the task offloading and the computation resource allocation are decided based on the proposed GDMSG.
       \item \textit{Equalizing resources, GDMSG-based offloading and phase shift (ERGOPS)}: the computation resource allocation of the BS are decided averagely, while the strategies of task offloading and IRS phase shift vector are determined based on the proposed GDMSG.
       \item \textit{DRL-based offloading, phase shift, and
       resource allocation (DOPSRA)}: the task offloading, IRS phase shift vector, and computation resource allocation strategies are decided through a PPO-based DRL approach.
\end{itemize}
\vspace{-2ex}
\subsection{Evaluation Results}
\label{sec_Evaluation Results}
\vspace{-0.2ex}
\par Figs. \ref{fig_time}(a), \ref{fig_time}(b), \ref{fig_time}(c), and \ref{fig_time}(d) illustrate the impact of time on total task completion delay, total energy consumption, QoE of vehicles, and revenue of BS. Clearly, the proposed GDMSG outperforms other approaches in minimizing total task completion delay and maximizing QoE of vehicles and revenue of BS. This is mainly because GDMSG leverages the GDM to capture the multidimensional characteristics of complex environments and dynamically optimizes the considered strategies. This reduces resource conflicts and communication bottlenecks, thereby enhancing task processing efficiency and ensuring low delay. Moreover, the strategy search capability of the GDM enables the system to identify globally optimal solutions in dynamic environments. Even under increased load, it ensures efficient resource allocation, prevents delays and shortages from affecting service quality, and boosts revenue of BS through effective resource management. It is worth noting that the reduction in delay comes with increased energy consumption, as GDMSG demands more frequent computation and resource scheduling to enhance task efficiency and service quality. However, the increase in energy consumption is justified because the significant improvements in QoE of vehicles and revenue of BS outweigh the additional costs. 

\par In comparison, the performance of DOPSRA and ERGOPS is inferior to that of GDMSG. This is primarily because ERGOPS does not optimize resource allocation, which results in low resource utilization and increased task processing delays. Although ERGOPS has lower energy consumption, its effectiveness in enhancing QoE of vehicle and revenue of BS is limited. Furthermore, DOPSRA relies on the PPO algorithm, whose local exploration strategy often struggles to quickly find global optimal solutions, thereby resulting in higher delays and impacting system performance. Additionally, RPSGORA and ROPSRA perform poorly across all metrics, primarily due to their inherent randomness, which causes instability in environments with high resource competition or heavy loads.

\section{Conclusion}
\label{sec_conclusion}
\par In this work, we \textcolor{b}{studied} the multi-objective optimization of task offloading, IRS phase shift vector, and computation resource allocation in IRS-assisted MEC \textcolor{b}{systems} for vehicular networks. We first \textcolor{b}{formulated} the multi-objective optimization problem with the aim of minimizing the total task completion delay and total energy consumption of the system. Then, we \textcolor{b}{proposed} the GDMSG to solve the problem, which is reformulated \textcolor{b}{into} a Stackelberg game framework and uses GDM to solve the reformulated problem at a lower cost. Simulation results \textcolor{b}{indicated} that the proposed GDMSG illustrates outstanding performance with respect to the total task completion delay, QoE of vehicles, and revenue of BS.

\section*{Acknowledgement}
\par This study is supported in part by the National Natural Science Foundation of China (62172186, 62272194, 62471200), in part by the Science and Technology Development Plan Project of Jilin Province (20230201087GX), in part by the China Postdoctoral Science Foundation General Fund (2023M731282), and in part by the U.S. National Science Foundation (NSF) (CNS-2415208).

\ifCLASSOPTIONcaptionsoff
\newpage
\fi
\bibliographystyle{IEEEtran}
\bibliography{references.bib}

\begin{thebibliography}{10}
\providecommand{\url}[1]{#1}
\csname url@samestyle\endcsname
\providecommand{\newblock}{\relax}
\providecommand{\bibinfo}[2]{#2}
\providecommand{\BIBentrySTDinterwordspacing}{\spaceskip=0pt\relax}
\providecommand{\BIBentryALTinterwordstretchfactor}{4}
\providecommand{\BIBentryALTinterwordspacing}{\spaceskip=\fontdimen2\font plus
\BIBentryALTinterwordstretchfactor\fontdimen3\font minus \fontdimen4\font\relax}
\providecommand{\BIBforeignlanguage}[2]{{%
\expandafter\ifx\csname l@#1\endcsname\relax
\typeout{** WARNING: IEEEtran.bst: No hyphenation pattern has been}%
\typeout{** loaded for the language `#1'. Using the pattern for}%
\typeout{** the default language instead.}%
\else
\language=\csname l@#1\endcsname
\fi
#2}}
\providecommand{\BIBdecl}{\relax}
\BIBdecl

\bibitem{Fu2022}
Y.~Fu, C.~Li, F.~R. Yu, T.~H. Luan, and Y.~Zhang, ``A survey of driving safety with sensing, vehicular communications, and artificial intelligence-based collision avoidance,'' \emph{{IEEE} Trans. Intell. Transp. Syst.}, vol.~23, no.~7, pp. 6142--6163, 2022.

\bibitem{Wang2019}
J.~Wang, J.~Liu, and N.~Kato, ``Networking and communications in autonomous driving: {A} survey,'' \emph{{IEEE} Commun. Surv. Tutorials}, vol.~21, no.~2, pp. 1243--1274, 2019.

\bibitem{Porambage2018}
P.~Porambage, J.~Okwuibe, M.~Liyanage, M.~Ylianttila, and T.~Taleb, ``Survey on multi-access edge computing for internet of things realization,'' \emph{{IEEE} Commun. Surv. Tutorials}, vol.~20, no.~4, pp. 2961--2991, 2018.

\bibitem{Huang2023}
Z.~Huang, B.~Zheng, and R.~Zhang, ``Roadside irs-aided vehicular communication: Efficient channel estimation and low-complexity beamforming design,'' \emph{{IEEE} Trans. Wirel. Commun.}, vol.~22, no.~9, pp. 5976--5989, 2023.

\bibitem{Li2024}
Y.~Li, L.~Li, and P.~Fan, ``Mobility-aware computation offloading and resource allocation for {NOMA} {MEC} in vehicular networks,'' \emph{{IEEE} Trans. Veh. Technol.}, vol.~73, no.~8, pp. 11\,934--11\,948, 2024.

\bibitem{Zhao2023}
S.~Zhao, Y.~Liu, S.~Gong, B.~Gu, R.~Fan, and B.~Lyu, ``Computation offloading and beamforming optimization for energy minimization in wireless-powered irs-assisted {MEC},'' \emph{{IEEE} Internet Things J.}, vol.~10, no.~22, pp. 19\,466--19\,478, 2023.

\bibitem{Wan2024}
Z.~Wan, W.~Jiang, J.~Nie, D.~Niyato, C.~Pan, and Z.~Xiong, ``Min-max fairness based joint optimal design for {IRS}-assisted {MEC} systems,'' \emph{{IEEE} Trans. Veh. Technol.}, vol.~73, no.~8, pp. 11\,949--11\,963, 2024.

\bibitem{Jiang2023}
F.~Jiang, Y.~Peng, K.~Wang, L.~Dong, and K.~Yang, ``{MARS:} {A} drl-based multi-task resource scheduling framework for {UAV} with irs-assisted mobile edge computing system,'' \emph{{IEEE} Trans. Cloud Comput.}, vol.~11, no.~4, pp. 3700--3712, 2023.

\bibitem{Cao2024}
H.~Cao, C.~Tan, Z.~Gao, Y.~Xu, G.~Chen, P.~Heng, and S.~Z. Li, ``A survey on generative diffusion models,'' \emph{{IEEE} Trans. Knowl. Data Eng.}, vol.~36, no.~7, pp. 2814--2830, 2024.

\bibitem{Ali2016}
M.~S. Ali, H.~Tabassum, and E.~Hossain, ``Dynamic user clustering and power allocation for uplink and downlink non-orthogonal multiple access {(NOMA)} systems,'' \emph{{IEEE} Access}, vol.~4, pp. 6325--6343, 2016.

\bibitem{Zhang2021}
H.~Zhang, B.~Di, Z.~Han, H.~V. Poor, and L.~Song, ``Reconfigurable intelligent surface assisted multi-user communications: {H}ow many reflective elements do we need?'' \emph{{IEEE} Wirel. Commun. Lett.}, vol.~10, no.~5, pp. 1098--1102, 2021.

\bibitem{Zhang2017}
J.~Zhang, L.~Dai, Z.~He, S.~Jin, and X.~Li, ``Performance analysis of mixed-adc massive {MIMO} systems over rician fading channels,'' \emph{{IEEE} J. Sel. Areas Commun.}, vol.~35, no.~6, pp. 1327--1338, 2017.

\bibitem{Di2020}
B.~Di, H.~Zhang, L.~Song, Y.~Li, Z.~Han, and H.~V. Poor, ``Hybrid beamforming for reconfigurable intelligent surface based multi-user communications: Achievable rates with limited discrete phase shifts,'' \emph{{IEEE} J. Select. Areas Commun.}, vol.~38, no.~8, pp. 1809--1822, 2020.

\bibitem{Tong2023}
S.~Tong, Y.~Liu, J.~V. Misic, X.~Chang, Z.~Zhang, and C.~Wang, ``Joint task offloading and resource allocation for fog-based intelligent transportation systems: {A} {UAV}-enabled multi-hop collaboration paradigm,'' \emph{{IEEE} Trans. Intell. Transp. Syst.}, vol.~24, no.~11, pp. 12\,933--12\,948, 2023.

\bibitem{ho2020denoising}
J.~Ho, A.~Jain, and P.~Abbeel, ``Denoising diffusion probabilistic models,'' \emph{Adv. Neural Inf. Process. Syst.}, vol.~33, pp. 6840--6851, 2020.

\end{thebibliography}

\end{document}